\documentclass[12pt]{cernart}
\usepackage{times}
\usepackage{epsfig}
\usepackage{amssymb}
\usepackage{cite}
\usepackage{colordvi}
\usepackage{pstricks}

\newcommand{\GeV}{\mathrm{GeV}}

\begin {document}

\begin{titlepage}
\docnum{%
\vbox{CERN--PH--EP/2006--029\\
      v2 21 November 2006}
}
\date{v1 21 September 2006}

%\vspace*{2.5cm}
\vspace{1cm}

\begin{center}
{\LARGE {\bf The Deuteron Spin-dependent Structure \\ Function $g_1^d$
         and its First Moment\\
              }}
\vspace*{0.5cm}
%\normalsize
\end{center}

\author{\large The COMPASS Collaboration}
\vspace{2cm}

\begin{abstract}
 We present a measurement of the deuteron spin-dependent structure
function $g_1^d$ based on the data collected by the COMPASS experiment at CERN
during the years 2002--2004. The data provide an accurate evaluation for
 $\Gamma_1^{d}$,
the first
moment of $g_1^d(x)$,  and  for the matrix element of the singlet axial current, $a_0$.
The results of QCD fits in the next to leading order (NLO) on all $g_1$
deep inelastic scattering data are also presented. They
provide two solutions with the gluon spin distribution function $\Delta G$
positive or negative,
%$\Delta G(x) > 0$ or $\Delta G(x) < 0$,
which describe the data equally well.
In both cases, at $Q^2 = 3(\GeV/c)^2$ the first moment of $\Delta G(x)$ is found to be
of the order of 0.2 -- 0.3 in absolute value.
\vfill
\submitted{(To be Submitted to Physics Letters B)}
\end{abstract}

\begin{Authlist}
{\large ~~~~~~~~~~~~~~~~~~~The COMPASS Collaboration} \newline \newline %\\[\baselineskip]
%%%%%%%%%%%%%%%%%%%%%%%%%%%%%%%%%%%%%%%%%%%%%%%%%%%%%%%%%%%%%%%%%%%%%%%%%%%%%%%%%%%%%%%%%%%%%%%%%%%%%%%%%%%%%%%%%%%%%%%
%
%   authors A1 paper - plb style             9-Feb-2005
%
%%%%%%%%%%%%%%%%%%%%%%%%%%%%%%%%%%%%%%%%%%%%%%%%%%%%%%%%%%%%%%%%%%%%%%%%%%%%%%%%%%%%%%%%%%%%%%%%%%%%%%%%%%%%%%%%%%%%%%%
%E.S.~Ageev\Iref{protvino},  				left end 2004
V.Yu.~Alexakhin\Iref{dubna},
Yu.~Alexandrov\Iref{moscowlpi},
G.D.~Alexeev\Iref{dubna},
M.~Alexeev\Iref{turin},
A.~Amoroso\Iref{turin},
B.~Bade{\l}ek\Iref{warsaw},
F.~Balestra\Iref{turin},
J.~Ball\Iref{saclay},
J.~Barth\Iref{bonnpi},
G.~Baum\Iref{bielefeld}
% ...
\samepage
\end{Authlist}
\samepage
%%%%%%%%%%%%%%%%%%%%%%%%%%%%%%%%%%%%%%%%%%%%%%%%%%%%%%%%%%%%%%%%%%%%%%%%%%%%%%%%%%%%%%%%%%%%%%%%%%%%%%%%%%%%%%%%%%%%%%%
%
% institutes
%
%%%%%%%%%%%%%%%%%%%%%%%%%%%%%%%%%%%%%%%%%%%%%%%%%%%%%%%%%%%%%%%%%%%%%%%%%%%%%%%%%%%%%%%%%%%%%%%%%%%%%%%%%%%%%%%%%%%%%%%
\Instfoot{bielefeld}{ Universit\"at Bielefeld, Fakult\"at f\"ur Physik, 33501 Bielefeld, Germany\Aref{g}}
\Instfoot{bochum}{ Universit\"at Bochum, Institut f\"ur Experimentalphysik, 44780 Bochum, Germany\Aref{g}}
\Instfoot{bonniskp}{ Universit\"at Bonn, Helmholtz-Institut f\"ur  Strahlen- und Kernphysik, 53115 Bonn, Germany\Aref{g}}
\Instfoot{bonnpi}{ Universit\"at Bonn, Physikalisches Institut, 53115 Bonn, Germany\Aref{g}}
\Instfoot{brno}{Institute of Scientific Instruments, AS CR, 61264 Brno, Czech Republic\Aref{h}}
\Instfoot{burdwan}{ Burdwan University, Burdwan 713104, India\Aref{j}}
\Instfoot{calcutta}{ Matrivani Institute of Experimental Research \& Education, Calcutta-700 030, India\Aref{k}}
\Instfoot{dubna}{ Joint Institute for Nuclear Research, 141980 Dubna, Moscow region, Russia}
\Instfoot{erlangen}{ Universit\"at Erlangen--N\"urnberg, Physikalisches Institut, 91054 Erlangen, Germany\Aref{g}}
\Instfoot{freiburg}{ Universit\"at Freiburg, Physikalisches Institut, 79104 Freiburg, Germany\Aref{g}}
\Instfoot{cern}{ CERN, 1211 Geneva 23, Switzerland}
\Instfoot{heidelberg}{ Universit\"at Heidelberg, Physikalisches Institut,  69120 Heidelberg, Germany\Aref{g}}
\Instfoot{helsinki}{ Helsinki University of Technology, Low Temperature Laboratory, 02015 HUT, Finland  and University of Helsinki, Helsinki Institute of  Physics, 00014 Helsinki, Finland}
\Instfoot{liberec}{Technical University in Liberec, 46117 Liberec, Czech Republic\Aref{h}}
\Instfoot{lisbon}{ LIP, 1000-149 Lisbon, Portugal\Aref{i}}
\Instfoot{mainz}{ Universit\"at Mainz, Institut f\"ur Kernphysik, 55099 Mainz, Germany\Aref{g}}
\Instfoot{miyazaki}{University of Miyazaki, Miyazaki 889-2192, Japan\Aref{l}}
\Instfoot{moscowlpi}{Lebedev Physical Institute, 119991 Moscow, Russia}
\Instfoot{munichlmu}{Ludwig-Maximilians-Universit\"at M\"unchen, Department f\"ur Physik, 80799 Munich, Germany\Aref{g}}
\Instfoot{munichtu}{Technische Universit\"at M\"unchen, Physik Department, 85748 Garching, Germany\Aref{g}}
\Instfoot{nagoya}{Nagoya University, 464 Nagoya, Japan\Aref{l}}
\Instfoot{praguecu}{Charles University, Faculty of Mathematics and Physics, 18000 Prague, Czech Republic\Aref{h}}
\Instfoot{praguectu}{Czech Technical University in Prague, 16636 Prague, Czech Republic\Aref{h}}
%\Instfoot{aCTUa}{Czech Technical University, Faculty of Nuclear Sciences and Physical Engineering, 11519 Prague, Czech Republic\Aref{h}}
%\Instfoot{aCTUb}{Czech Technical University, Faculty of Mechanical Engineering, 16636 Prague, Czech Republic\Aref{h}}
\Instfoot{protvino}{ State Research Center of the Russian Federation, Institute for High Energy Physics, 142281 Protvino, Russia}
\Instfoot{saclay}{ CEA DAPNIA/SPhN Saclay, 91191 Gif-sur-Yvette, France}
\Instfoot{telaviv}{ Tel Aviv University, School of Physics and Astronomy, 
              %Raymond and Beverly Sackler Faculty of Exact Sciences, 
              69978 Tel Aviv, Israel\Aref{m}}
\Instfoot{triestictp}{ INFN Trieste and ICTP--INFN MLab Laboratory, 34014 Trieste, Italy}
\Instfoot{triest}{ INFN Trieste and University of Trieste, Department of Physics, 34127 Trieste, Italy}
\Instfoot{turin}{ INFN Turin and University of Turin, Physics Department, 10125 Turin, Italy}
\Instfoot{warsaw}{ So{\l}tan Institute for Nuclear Studies and Warsaw University, 00-681 Warsaw, Poland\Aref{n} }
\Instfoot{warsawtu}{ Warsaw University of Technology, Institute of Radioelectronics, 00-665 Warsaw, Poland\Aref{o} }
\Anotfoot{a}{Also at IST, Universidade T\'ecnica de Lisboa, Lisbon, Portugal}
\Anotfoot{b}{Also at University of East Piedmont, 15100 Alessandria, Italy}
\Anotfoot{c}{deceased}
\Anotfoot{d}{On leave of absence from JINR Dubna}               
\Anotfoot{e}{Also at Chubu University, Kasugai, Aichi, 487-8501 Japan}
\Anotfoot{f}{Also at Yamagata University, Yamagata, 992-8510 Japan}
\Anotfoot{g}{Supported by the German Bundesministerium f\"ur Bildung und Forschung}
\Anotfoot{h}{Suppported by Czech Republic MEYS grants ME492 and LA242}
\Anotfoot{i}{Supported by the Portuguese FCT - Funda\c{c}\~ao para
               a Ci\^encia e Tecnologia grants POCTI/FNU/49501/2002 and POCTI/FNU/50192/2003}
\Anotfoot{j}{Supported by DST-FIST II grants, Govt. of India}
\Anotfoot{k}{Supported by  the Shailabala Biswas Education Trust}
\Anotfoot{l}{Supported by the Ministry of Education, Culture, Sports,
               Science and Technology, Japan; Daikou Foundation  and Yamada Foundation}
\Anotfoot{m}{Supported by the Israel Science Foundation, founded by the Israel Academy of Sciences and Humanities}
\Anotfoot{n}{Supported by KBN grant nr 621/E-78/SPUB-M/CERN/P-03/DZ 298 2000 and
             nr 621/E-78/SPB/CERN/P-03/DWM 576/2003--2006,
             and by MNII reseach funds for 2005--2007}
\Anotfoot{o}{Supported by  KBN grant nr 134/E-365/SPUB-M/CERN/P-03/DZ299/2000}

\vfill
\hbox to 0pt {}

\end{titlepage}

\noindent
The spin structure function $g_1^d$ of the deuteron has been measured
for the first time almost 15 years ago by the SMC experiment at CERN \cite{SMC_92}.
Since then, high accuracy measurements of $g_1^d$
in the deep inelastic scattering (DIS) region have been performed
at SLAC \cite{e143,E155_d} and DESY \cite{HERMES}.
Due to the relatively low incident energy, the %deep inelastic scattering
DIS events collected in {those} experiments
cover only a limited range of $x$ for $Q^2 > 1(\GeV/c)^2$,
$x > 0.015$ and $x> 0.03$, respectively.
Further measurements covering the low $x$ region were also performed at
CERN (see \cite{smc} and references therein).
Besides its general interest for the understanding of the spin structure
of the nucleon, $g_1^d$ is specially important because its first moment is directly related
to the
matrix element of the singlet axial vector current $a_0$.
A precise measurement of $g_1^d$ can thus
provide an evaluation of the fraction of nucleon spin carried by quarks, on the condition
that the covered range extends far enough to low $x$ to provide a reliable value of the
first moment.

Here we present new results from the COMPASS experiment at CERN on the deuteron spin
asymmetry $A_1^d$ and the spin-dependent structure function $g_1^d$ covering the range
$1(\GeV/c)^2 < Q^2 < 100(\GeV/c)^2$ in the photon virtuality and $0.004<x<0.7$ in
the Bjorken scaling variable. The data
sample used in the present analysis was  collected during the years 2002--2004
and corresponds to an integrated luminosity of about 2 fb$^{-1}$.  Partial results
based on the data collected during the first two years {of the data taking}
have been published in Ref.~\cite{cmp23}.
At the time, the values of $g_1^d$ were not precise enough, in particular at large $x$,
to allow a meaningful evaluation of the first moment, $\Gamma_1^{d}$. The results presented
here are based on a 2.5 times larger statistics and supersede those of Ref.~\cite{cmp23}.
We refer the reader to this reference for the description of  the 160~GeV muon beam,
the $^6$LiD polarised target and the COMPASS spectrometer which remained
basically unchanged in 2004.
A global fit to all $g_1^{p,n,d}$ data is needed to evolve the $g_1^d(x_i,Q^2_i)$
measurements to a common $Q^2$. As previous fits were found to be in disagreement with
our data at low $x$, we have performed a new QCD fit at NLO.
The resulting polarised parton distribution functions (PDF) are also presented in this paper and
discussed in relation with the new data, however without a full investigation of
the theoretical uncertainties due, for instance, to the values of the factorisation and
renormalisation scales.

\begin{figure}
  \begin{minipage}{0.49\linewidth}
\noindent
    \includegraphics[width=\textwidth,clip]{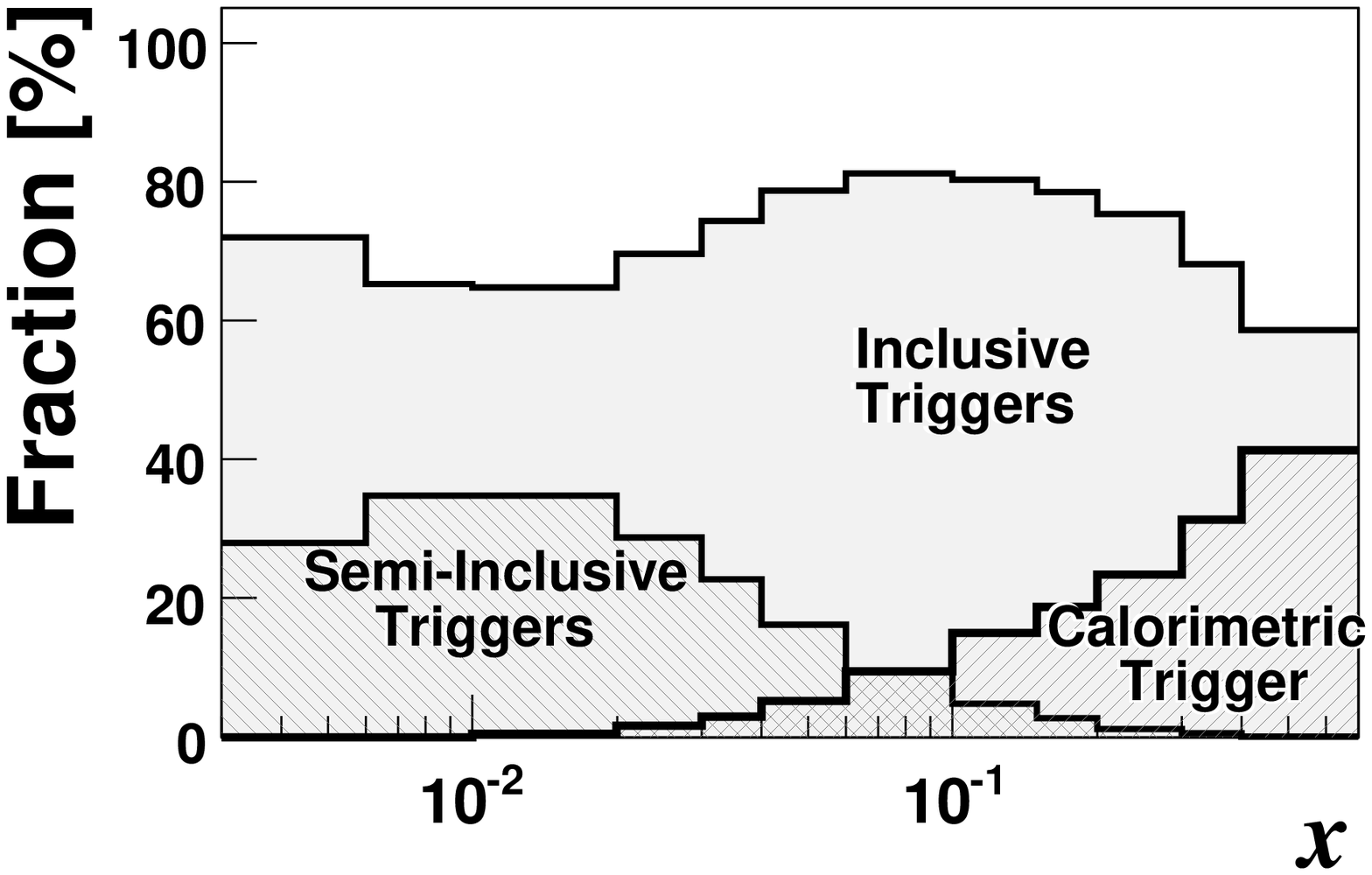}
    \caption{\small Fraction of inclusive, semi-inclusive, and calorimetric triggers
      as a function of $x$.
      Events are counted with the weight they carry in the asymmetry calculation.}
    \label{fig:trigg_xq2}
  \end{minipage}
  \hfill
  \begin{minipage}{0.49\linewidth}
\noindent
    \includegraphics[width=\textwidth,clip]{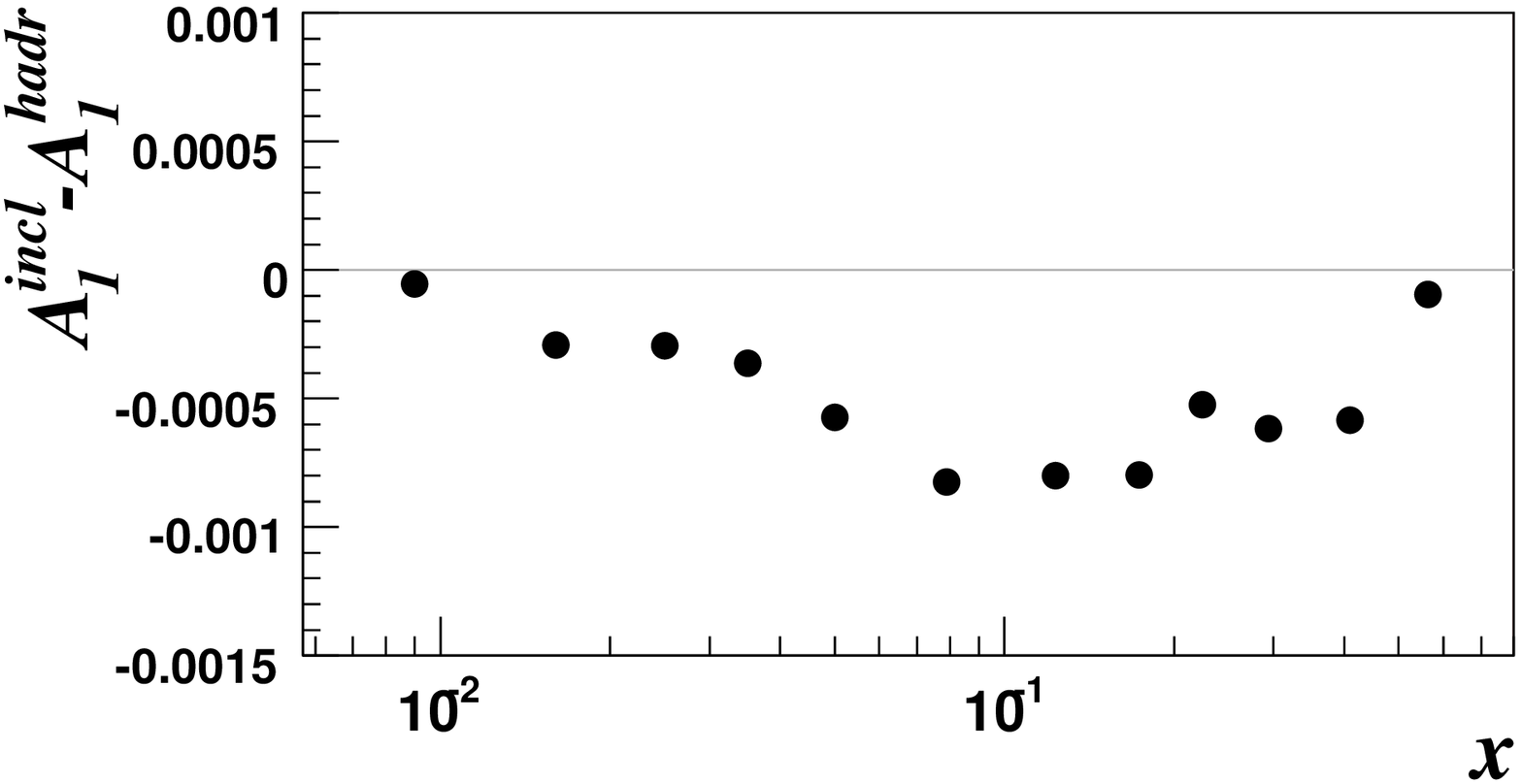}
    \caption{\small Difference between asymmetries for inclusive and hadronic
       Monte Carlo events in the kinematic range covered by the purely calorimetric trigger.}
    \label{fig:Bias}
  \end{minipage}
\end{figure}

The COMPASS data acquisition system
is triggered by coincidence signals in hodoscopes,
defining the direction of the scattered muon behind { the spectrometer magnets},
and by signals in the hadron calorimeters \cite{trigger}.
Triggers due to halo muons are eliminated
by veto counters installed upstream of the target.
Inclusive triggers, based on muon detection only, cover the full range of $x$ and are dominant
in the medium ($x$, $Q^2$) region. Semi-inclusive triggers, based on the muon energy
loss and the presence of a hadron signal in the calorimeters, contribute mainly at
low $x$ and low $Q^2$.  Purely calorimetric triggers, based on the energy deposit in the
hadron calorimeter without any condition on the scattered muon,
account for most events at large $Q^2$.
The relative contributions of these three trigger types are
shown in Fig.~\ref{fig:trigg_xq2}  as a function of $x$. % and $Q^2$.
The minimum hadron
energy deposit required for the purely calorimetric trigger has been reduced to 10~GeV
for the events collected in 2004. As a consequence, the contribution of this trigger
now reaches 40\% at large $x$, compared to 20\% in 2002--2003 (Ref.~\cite{cmp23}).

All events used in the present analysis require the presence of reconstructed beam muon
and scattered muon trajectories defining an interaction point, which is located inside
one of the target cells. The momentum of the incoming muon, measured in the beam
spectrometer, is centered around 160~GeV$/c$ with an RMS of 8~GeV$/c$
for the {Gaussian} core. In the present analysis its value
is required to be between 140 and 180~GeV$/c$. In addition
the extrapolated beam muon trajectory is required to cross entirely both target cells in order to
equalize the fluxes seen by  each of them.
 The scattered muon is identified by signals collected behind the hadron absorbers and
(except for the purely calorimetric trigger) its trajectory must be consistent with the
hodoscope signals defining the event trigger. For hadronic triggers, a second outgoing
reconstructed track is required at  the interaction point.
%The \Blue{deep inelastic scattering} {DIS} events \Blue{(DIS)} used in the present analysis
The DIS events used in the present analysis
are selected by cuts on the four-momentum transfer
squared ($Q^2 > 1(\GeV/c)^2$) and the fractional energy of the virtual photon ($0.1 < y < 0.9$).
The resulting sample consists of $89\times 10^6$ events, out of which about 10\% were obtained in 2002,
30\% in 2003 and 60\% in 2004.
In order to extend  the coverage of the low $x$ region, we also analyse events
in the interval $0.003 < x < 0.004$
selected in the same way but with a  $Q^2$ cut lowered to 0.7~(GeV$/c)^2$.
These events are included in the figures but not used in QCD
calculations or moment estimation, in view of their low $Q^2$.

During  data taking the two target cells are polarised in opposite directions, so that
the deuteron spins are parallel (${\uparrow \uparrow}$) or antiparallel
(${\uparrow \downarrow}$) to the spins of the incoming muons. The spins are inverted
every 8 hours by a rotation of the target magnetic field.
The average beam and target polarisations are about
$-0.80$ ($-0.76$ in 2002 and 2003) and $\pm 0.50$, respectively.

The cross-section asymmetry $A^d = (\sigma^{\uparrow \downarrow} - \sigma^{\uparrow \uparrow}) /
(\sigma^{\uparrow \downarrow} + \sigma^{\uparrow \uparrow})$, for antiparallel
($\uparrow \downarrow$) and parallel ($\uparrow \uparrow$)  spins of the incoming muon
and the target deuteron can be obtained from the numbers of events $N_i$ % (i=1,4)$
collected from each cell before and
after reversal of the target spins:
\begin{equation}
N_i = a_i \phi_i n_i {\overline \sigma} (1 + P_B P_T f A^d),
~~~~~i=1,2,3,4,
\label{eq:CountRate}
\end{equation}
where $a_i$ is the acceptance, $\phi_i$ the incoming flux, $n_i$ the number of target nucleons,
${\overline \sigma}$ the spin-averaged cross-section, $P_B$ and $P_T$ the beam and target
polarisations and $f$ the target dilution factor.
The latter includes a corrective factor $\rho=\sigma_d^{1\gamma}/\sigma_d^{tot}$ \cite{terad}
accounting for radiative events on the {unpolarised} deuteron
and a correction for the relative polarisation of deuterons bound in $^6$Li
compared to free deuterons.
Fluxes and acceptances cancel out in the asymmetry calculation on the condition
that the ratio of the acceptances of the two cells is the same before and after spin
reversal \cite{smc_97}.

The longitudinal virtual-photon deuteron asymmetry, $A_1^d$, is defined
{via the asymmetry of absorption cross-sections of transversely polarised photons}
as
\begin{equation}
A_1^d = (\sigma_0^T - \sigma_2^T) / (2 \sigma^T),
\label{eq:AsymDeuteron}
\end{equation}
where $\sigma_J^T$ is the $\gamma^{*}$-deuteron absorption cross-section for a total
spin projection $J$ and
$\sigma^T$ is the total transverse photoabsorption cross-section. % of transverse $\gamma^{*}$.}
The relation between $A_1^d$ and the experimentally measured $A^d$  is %reduces to
\begin{equation}
%A^d_1 = A^d/D - \eta A^d_1 \simeq  A^d/D ,
A^d = D(A^d_1 + \eta A^d_2) ,
\label{eq:A1d_1}
\end{equation}
where $D$ and $\eta$ depend on kinematics.
The transverse asymmetry $A_2^d$ has been measured at SLAC and
 found to be small \cite{e155_a2}.
In view of this, in our analysis, Eq.~(\ref{eq:A1d_1}) has been reduced to $A^d_1\simeq A^d/D$.
The virtual-photon depolarisation factor $D$ depends on %which depends on the event kinematics and on
the %unpolarised
%spin-independent function $R = \sigma^L/\sigma^T$.
{ratio of longitudinal and transverse photoabsorption cross sections $R = \sigma^L/\sigma^T$}.
In the present analysis an
updated parametrisation of $R$ taking into account all existing
measurements is  used  \cite{R_1998}.
The tensor-polarised structure function of the deuteron
has been measured by HERMES \cite{hermes_b1} and its effect
on the measurement of the longitudinal spin structure was found to be negligible, which
justifies the use of Eqs~(\ref{eq:CountRate}--\ref{eq:A1d_1}) in the present analysis.

In order to minimize the statistical error of the asymmetry, the kinematic factors
$f$, $D$ and {the beam polarisation} $P_B$  are calculated event-by-event and
used to weight events.
A parametrisation of $P_B$ as a function of the beam momentum is used,
while for $P_T$ an average value is used for the data sample
taken between two consecutive target spin reversals.
% {over about 30 minutes of run time} value is taken. % for each run.
The obtained asymmetry is corrected for spin-dependent
radiative effects  according to Ref.~\cite{polrad}.
The asymmetry  is evaluated separately for inclusive and for hadronic events
because the dilution factors and the radiative corrections to the asymmetry are different.
This is because the correction due to radiative elastic  and quasi-elastic scattering
events  only affects the inclusive sample.

It has been checked that the use of hadronic triggers does not bias the inclusive
asymmetries. The most critical case is for the calorimetric trigger events at large $x$, where
high-energy hadron production is limited by kinematics. This effect has been studied by
Monte Carlo, using the program POLDIS \cite{poldis}. DIS events were generated within the
acceptance of the calorimetric trigger and their asymmetry calculated analytically at
the leading order. A selection based on the hadron requirements corresponding to the
trigger was applied and the asymmetries for the selected sample  compared to
the original ones. The differences were found to be smaller than 0.001 in all intervals
of $x$ (Fig.~\ref{fig:Bias}) and thus  negligible, so that inclusive and hadronic
asymmetries can be safely combined for further analysis
%A similar conclusion was reached in
{(see also
the SMC analysis \cite{smc})}.

\begin{figure}
\begin{center}
\epsfig{file=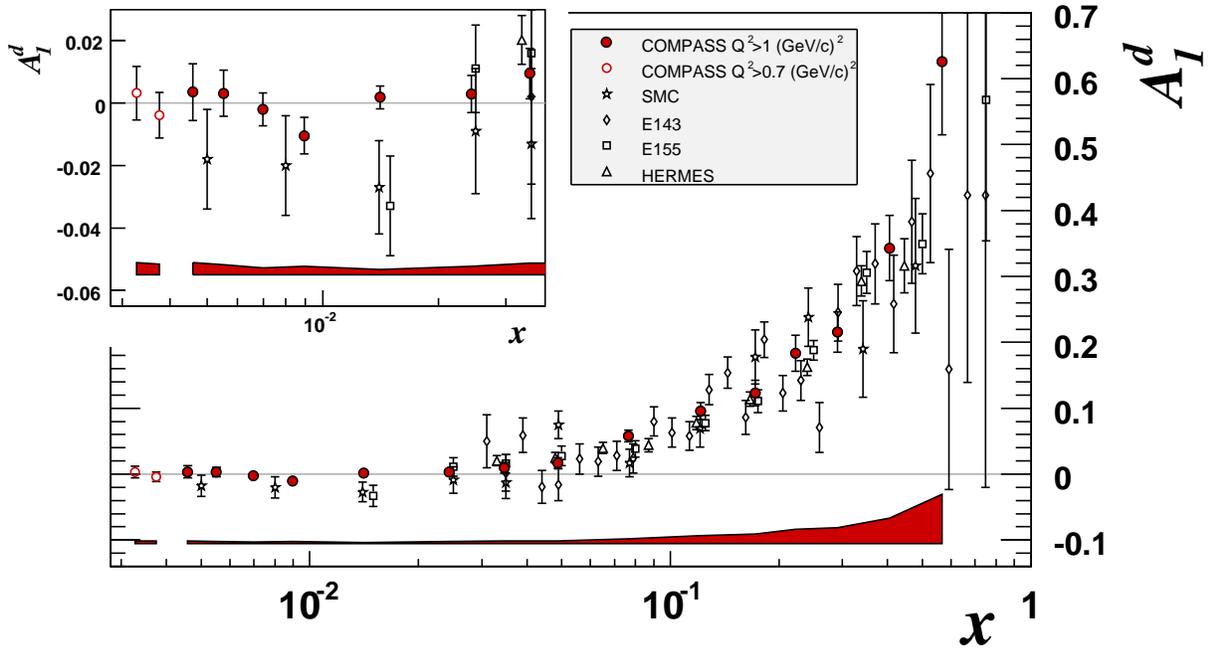,width=\textwidth}
\end{center}
\caption{\small The asymmetry $A_1^d(x)$ as  measured in COMPASS and previous results from SMC
  \cite{smc}, HERMES \cite{HERMES},  SLAC E143 \cite{e143} and E155 \cite{E155_d}
  at  $Q^2>1(\GeV/c)^2$. The SLAC values of $g_1/F_1$ have been converted to $A_1$
  and the E155 data corresponding to the same $x$ have been averaged over $Q^2$.
  Only statistical errors are shown with the data points.
  The shaded areas  show the size of the COMPASS systematic errors.}
\label{fig:A1_CMP_SMC}
\end{figure}

The final values of $A_1^d(x,Q^2)$, obtained as weighted averages of the asymmetries
in the inclusive and hadronic data sets, are listed in Table~\ref{tab:a1_g1} with the
corresponding average values of $x$ and $Q^2$.
%The values of $A_1^d$
They are also shown as a function of $x$ in Fig.~\ref{fig:A1_CMP_SMC}
in comparison with previous results from experiments at CERN \cite{smc},
DESY \cite{HERMES} and SLAC \cite{e143,E155_d}.
%They confirm
The values of $A_1^d$ confirm, with increased
statistical precision, the observation made in Ref.~\cite{cmp23} that the asymmetry
is consistent with zero for $x < 0.03$.
Values of $A_1^d$  originating from experiments at
different energies tend to coincide due to the very small $Q^2$ dependence of
$A_1^d$ at fixed $x$.

The systematic error of $A_1^d$ contains  multiplicative factors resulting from
uncertainties on $P_B$ and $P_T$, on the dilution factor $f$ and on the ratio
$R = \sigma^L/\sigma^T$ used to calculate the depolarisation factor $D$. When
combined in quadrature, these errors result in a global scale uncertainty of
10\% (Table~\ref{tab:sys_error}). The other important contribution to the systematic
error is due to false asymmetries which could be generated by instabilities in
some components of the spectrometer.
In order to minimize their effect, the values of $A_1^d$ in each interval of
$x$ have been calculated for 184 subsamples, each of them covering a short
period of running time and, therefore, ensuring similar detector operating
conditions.
An upper limit of the effect of detector instabilities has been
evaluated by a statistical approach.
The dispersion of the values of $A_1^d$ around their mean agrees with
the statistical error. There is thus no evidence for any broadening due to
time dependent effects.
%At 95\% confidence level,
Allowing the dispersion of $A_1^d$ to vary within its two standard deviations
we obtain an upper limit for the systematic error of $A_1^d$
in terms of its statistical precision: $\sigma_{syst}<0.4 \sigma_{stat}$.
%where $\sigma_{stat}$ denotes a statistical precision of the asymmetry.
% which is due to time variations of spectrometer components.
This estimation accounts for the time variation effects of spectrometer components.

Several other searches for false asymmetries were performed. Data from the
two target cells were combined in {different ways} in order to eliminate the
physical asymmetry. Data obtained with different settings of the microwave
frequencies, used to polarise the target by dynamic nuclear polarisation,
were compared. No evidence was found for any significant apparatus induced asymmetry.

\begin{figure}
\begin{center}
\epsfig{file=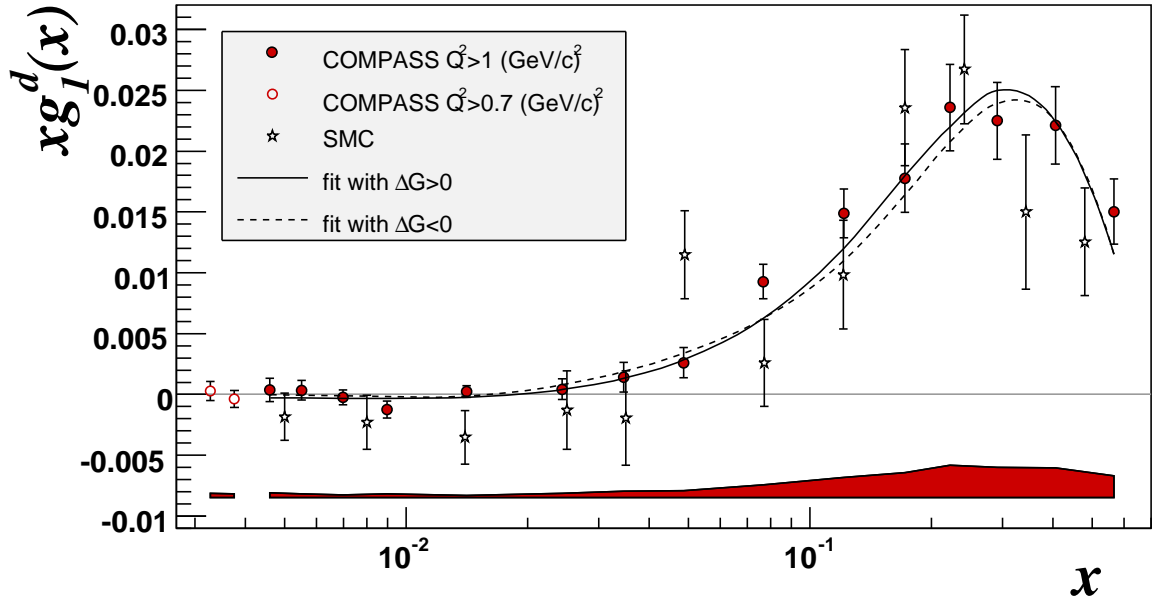,width=\textwidth}
\end{center}
\caption{Values of  $xg_1^d(x)$. % {\it vs}. $x$.
  The COMPASS points are given at the $\langle Q^2 \rangle$ where they were measured.
  The SMC points have been moved to the $Q^2$ of the corresponding COMPASS points.
  Only statistical errors are shown with the data points.
  The shaded band at the bottom shows the COMPASS systematic error.
  The curves show the results of QCD fits with $\Delta G > 0$ and $\Delta G < 0$.}
\label{fig:g1_CMP_SMC}
\end{figure}

The longitudinal spin structure function is obtained as
\begin{equation}
g_1^d = \frac{F_2^d}{2~ x~(1 + R)} A_1^d\,,
\end{equation}
where $F_2^d$ is the spin-independent deuteron structure function. The values of
$g_1^d$
listed in the last column of Table~\ref{tab:a1_g1}
have been calculated with the $F_2^d$ parametrisation of Ref.~\cite{smc},
which covers the range of our data,
and the new parametrisation of $R$ already used in the depolarisation factor.
The systematic errors on $g_1^d$ are obtained in the same way as for $A_1^d$,
with an  additional contribution from the uncertainty on $F_2^d$.
%and a partial cancellation of the error on $R$ between the factors $(1+R)$ and $D$.
%Figure \ref{fig:g1_CMP_SMC} shows
The values of $x$, $g_1^d(x)$ for the
COMPASS data and, for comparison, the SMC results \cite{smc} moved to the  $Q^2$
of the corresponding COMPASS point {are shown in Fig.~\ref{fig:g1_CMP_SMC}}.
The two curves on {the figure represent} the results of two QCD fits
at NLO, described below, at the measured $Q^2$ of each data point.

\begin{figure}
\begin{center}
\epsfig{file=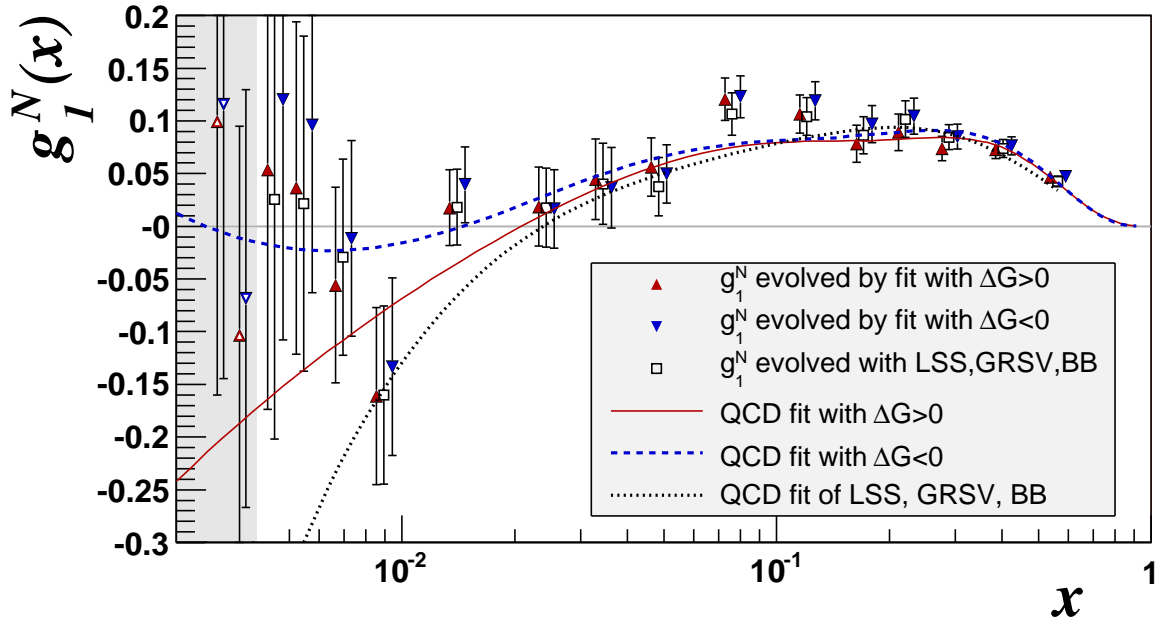,width=\textwidth}
\end{center}
\caption{\small The COMPASS values of $g_1^N$ evolved to $Q^2=3(\GeV/c)^2$.
  The open triangles at low $x$ correspond to $Q^2>0.7(\GeV/c)^2$, the other
  symbols to $Q^2>1(\GeV/c)^2$.
  Results of QCD fits are shown by curves. In addition to
  our fits ($\Delta G > 0$ and $\Delta G < 0$) the curve obtained with
  three {published polarised PDF parameterizations} (Bl\"{u}mlein and B\"{o}ttcher,
  GRSV and LSS05) \cite{Durham} is shown. These parameterizations lead almost to the
  same values of $g_1^N\left(x, Q^2=3(\GeV/c)^2\right)$ and  have been averaged.
  For clarity the data points evolved with different fits
  are shifted in $x$ with respect to each other.
  Only statistical errors are shown.}
\label{fig:g1_QCD_CMP_Q3}
%\vspace{0.6cm}
\end{figure}

The evaluation of the first moment $\Gamma_1^d(Q^2) = \int_0^1 g_1^d(x,Q^2) dx $
requires the evolution of all $g_1$ measurements to a common $Q_0^2$.
This is  done by using a fitted parametrisation
%of $g_1(x,Q^2)$,
$g_1^{fit}(x,Q^2)$, so that
\begin{equation}
%g_1(x_i,Q_0^2) = g_1(x_i,Q^2_i) + \Bigl [g_1^{fit}(x_i,Q_0^2) - g_1^{fit}(x_i,Q_i^2) \Bigr ].
g_1(x,Q_0^2) = g_1(x,Q^2) + \Bigl [g_1^{fit}(x,Q_0^2) - g_1^{fit}(x,Q^2) \Bigr ].
\end{equation}
We have used several fits of $g_1$ from the Durham data base \cite{Durham}:
 Bl\"{u}mlein-B\"{o}ttcher \cite{BB},
GRSV \cite{GRSV} and LSS05 \cite{LSS05}, and {we have} chosen $Q^2_0 = 3(\GeV/c)^2$
as reference $Q^2$ because it is close to the average $Q^2$ of the COMPASS DIS data.
The three parametrisations are quite similar in the
range of the COMPASS data and have been averaged. The resulting values of
$g_1^N = (g_1^p + g_1^n)/2$ are shown as open squares in Fig.~\ref{fig:g1_QCD_CMP_Q3}.
For clarity we now use $g_1^N$ instead of $g_1^d$ because the correction for the
D-wave state of the deuteron has been applied:
\begin{equation}
g_1^N(x,Q^2) = g_1^d(x,Q^2) / (1 - 1.5 \omega_D)
\end{equation}
with $\omega_D = 0.05 \pm 0.01$ \cite{omega_d}.
It can also be seen in Fig.~\ref{fig:g1_QCD_CMP_Q3} that the curve representing the
average of the three fits does not reproduce the trend of our data for $x < 0.02$ and
therefore cannot be used to estimate the unmeasured part of $g_1^N$ at low $x$.

In view of this, we have performed a new NLO {QCD} fit of all $g_1$ data
at $Q^2 > 1$ (GeV$/c)^2$ from proton,
deuteron and $^3$He targets, including the COMPASS data.
The deuteron data are from Refs.~\cite{smc,e143,E155_d,HERMES},
the proton data from Refs.~\cite{emc,smc,e143,e155p,HERMES} and
the $^3$He data from Refs.~\cite{e142,e154,jlab,hrm1}.

\begin{figure}
\begin{center}
\epsfig{file=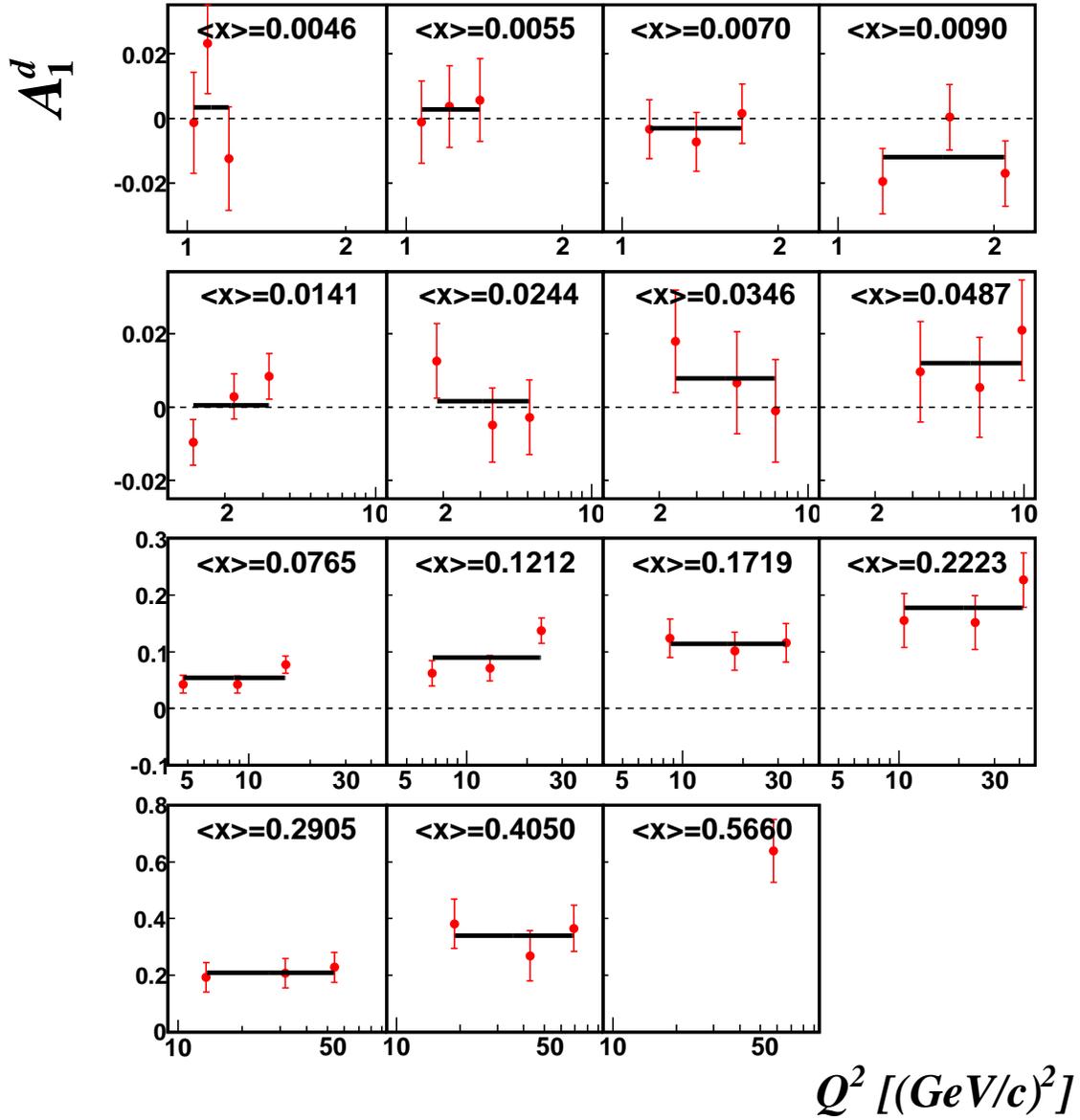,width=\textwidth}
\end{center}
\caption{\small Values of $A_1^d$ as a function of $Q^2$ in intervals of $x$.
     The solid lines show the results of fits to a constant.}
\label{A1_vs_Q2}
\end{figure}

In order to optimise the use of the COMPASS data in this fit,
all $x$ bins of Table~\ref{tab:a1_g1}, except the last one, have been subdivided
into three $Q^2$ intervals (Fig.~\ref{A1_vs_Q2}).
The number of COMPASS data points used in the fit is thus 43, out of a total of 230.

The fit is performed in the $\overline{\rm MS}$ renormalisation and
factorisation scheme and requires parametrisations of the quark singlet
spin distribution $\Delta \Sigma(x)$,
non-singlet distributions $\Delta q_3(x)$, $\Delta q_8(x)$
and the gluon spin distribution $\Delta G(x)$.
These distributions are given as an input at a reference $Q^2$ ($=Q^2_0$)
which is set to 3~(GeV$/c)^2$ and evolved according to the DGLAP equations.
The resulting values of $g_1(x,Q^2)$ are calculated for the $(x_i, Q_i^2)$
of each data point and compared to the experimental values.

The input parametrisations are written as
\begin{equation}
\Delta F_k = \eta_k  \frac{ x^{\alpha_k} \,(1-x)^{\beta_k} \,(1+\gamma_k x)}{\int_{0}^{1} x^{\alpha_k} \,(1-x)^{\beta_k} \,(1+\gamma_k x) dx} \,,
\end{equation}
where $\Delta F_k$ represents each of the polarised parton distribution
functions $\Delta \Sigma$, $\Delta  q_3$, $\Delta  q_8$ and $\Delta G$,
and $\eta_k$ is the integral of  $ \Delta F_k$.
The moments, $\eta_k$, of  the non-singlet distributions $\Delta q_3$ and $\Delta q_8$
are fixed by the baryon decay constants ($\rm F$$+$$\rm D$)  and ($\rm3F$$-$$\rm D$) respectively,
%\cite{a8}, assuming SU(3)$_f$ flavour symmetry.
assuming SU(3)$_f$ flavour symmetry.
The linear term $\gamma x$ is used only for the singlet distribution,
in which case the exponent $\beta_G$ is fixed because it is poorly constrained by the data.
% is also  fixed in \Green{some} of the fits.
This leaves 10 parameters in the input distributions.
In addition, the normalisation of E155 proton data is allowed to vary
within the limits quoted by the authors of Ref.~\cite{e155p}.

The optimal values of the parameters are obtained by minimizing the sum
\begin{equation}
\chi^2 = \sum_{i=1}^{{N=230}} \frac {\Bigl [ g_1^{{fit}}(x_i,Q^2_i) - g_1^{exp}(x_i,Q^2_i) \Bigr ]^2}
{\Bigl [ \sigma(x_i,Q^2_i)  \Bigr]^2} \,.
\end{equation}
Here the errors $\sigma$ are the statistical ones for all data sets, except for the proton data of
E155 where the uncorrelated part of the
systematic error on each point is added in quadrature to the
statistical one.
In order to keep the parameters in their physical range, the polarised strange sea
distribution
 $\Delta s(x) + \Delta {\overline s(x)}
= (1/3) (\Delta \Sigma(x) - \Delta q_8(x))$ is calculated
at every step and required to satisfy the positivity condition $\mid \Delta s(x) \mid \leq s(x)$
at all $Q^2$ {values}.  A similar condition is imposed on the gluon spin distribution $\Delta G(x)$.
The unpolarised distributions $s(x)$ and $G(x)$ used in this test are taken from the
MRST parametrisation \cite{MRST}.
This procedure leads to asymmetric errors on the parameters
when the fitted value is close to the allowed limit.

The fits have been performed with two different programs: the first
one uses the DGLAP evolution equations for the spin structure functions
\cite{smc_qcd}, the other one, referred to in \cite{Sissakian}, uses
the evolution of moments.
%Both
%programs give consistent values of the fitted PDFs
%and similar $\chi^2$-probabilities.
% an indication that the $Q^2$ evolution at NLO is properly described in both of them.
The fitted PDF parameters are compatible within one standard deviation and the
two programs give the same $\chi^2$-probabilities.
%Each program yields two solutions,
In each program the $\chi^2$ minimisation converges to two different solutions,
depending on the sign of the initial value of the gluon first moment $\eta_G$:
one solution with $\Delta G >0$,
the other one with $\Delta G < 0$ (Fig.~\ref{fig:g1_QCD_CMP_Q3}).
The fitted distributions of $g_1^N(x)$ differ at low $x$ but are both compatible with the data.
The two additional data points at $x < 0.004$ and $Q^2 > 0.7(\GeV/c)^2$,
not used in the fit, have  too large statistical errors to provide
a discrimination between the two solutions.
The  values of the parameters obtained in the  fits with positive and negative
$\Delta G$ are listed
in Table~\ref{tab:qcd_fit} with their statistical errors and will be discussed below.

The integral of $g_1^N$ in the measured region is obtained from the experimental values evolved
to a fixed $Q^2$ and averaged over the two fits.
{Taking into account} the contributions from the fits in the unmeasured
regions at low and high $x$ we obtain (Table~\ref{tab:Gamma_1}):
\begin{equation}
\Gamma_1^N\left(Q^2=3(\GeV/c)^2\right) = 0.050 \pm 0.003\ \mbox{(stat.)} \pm  0.003\ \mbox{(evol.)} \pm 0.005\ \mbox{(syst.)}.
\end{equation}
The second error accounts for the difference in $Q^2$ evolution between the two fits. The
systematic error is the dominant one and mainly corresponds to the 10\% scale uncertainty
resulting from the errors on the beam and target polarisations and on the dilution factor.

For comparison, the SMC result \cite{smc} was 
\begin{equation}
\Gamma_{1\mbox{, \sc smc}}^{N}\left(Q^2=10(\GeV/c)^2\right) = 0.021 \pm 0.007\ \mbox{(stat.)} \pm  0.014\ \mbox{(evol.)} \pm 0.003\ \mbox{(syst.)}.
\end{equation}
while our result evolved to $Q^2 = 10(\GeV/c)^2$ is
$0.051 \pm 0.003\ \mbox{(stat.)} \pm 0.003\ \mbox{(evol.)} \pm 0.005\ \mbox{(syst.)}$.
The difference between these two results reflects the fact
that the COMPASS data do not support the fast decrease of
$g_1^d\left(x,Q_0^2=3(\GeV/c)^2\right)$ at low $x$ which
was assumed in the SMC analysis, and thus force the fit to be different.
In the COMPASS analysis,
the  part of $\Gamma_1^N$ obtained from the measured region  represents 98\% of the total value.
This correction of only
2\% has to be compared to  a  correction  of about  50\% with respect to
the measured value in case of the SMC analysis \cite{smc}.

$\Gamma_1^N$ is of special interest because it gives access to
the  matrix element of the singlet axial current $a_0$
% $\hat{a}_0$ \cite{Larin},
which, except for a possible  gluon
contribution, measures the quark spin contribution to the nucleon spin.
At NLO, the relation between $\Gamma_1^N$ and $a_0$ reduces to
\begin{equation}
\Gamma_1^N(Q^2) = \frac{1}{9} \Bigl ( 1 - \frac{\alpha_s(Q^2)}{\pi}
+ {\cal O}(\alpha_s^2) \Bigr ) \Bigl ( a_0(Q^2) + \frac{1}{4} a_8 \Bigr ).
\end{equation}

From the COMPASS result on $\Gamma_1^N$ (Eq.~(9)) and taking
the value of $a_8$  from hyperon $\beta$ decay, assuming $SU(3)_f$ flavour
symmetry ($a_8 = 0.585\pm0.025$ \cite{a8}), one obtains
with the value of $\alpha_s$ evolved from the PDG value
$\alpha_s(m_Z^2) = 0.1187 \pm 0.005$
and assuming three active quark flavours:
\begin{equation}
a_0\left(Q^2=3(\GeV/c)^2\right) = 0.35 \pm 0.03\ \mbox{(stat.)} \pm 0.05\ \mbox{(syst.)}.
\end{equation}
The quoted systematic error accounts for the error from the evolution and for
the experimental systematic error, combined in quadrature.

The relation between $\Gamma_1^N$ and $a_0$ can also be rewritten in order to extract
the value of the matrix element $a_0$ in the limit
$Q^2 \rightarrow \infty$. % :
{Here we will follow a notation of Ref.~\cite{Larin} introducing a ``hat''
for the coefficient $C^S_1$ and $a_0$ at this limit}:
\[
\displaystyle \Gamma_1^N(Q^2) = \displaystyle
\frac{1}{9} ~\hat{C}^{S}_1(Q^2) ~ \hat{a}_0 ~+~ \frac{1}{36} ~ C^{NS}_1(Q^2) ~ a_8.
\]
The coefficients $\hat{C}_1^S$ and $C_1^{NS}$ have been calculated
in perturbative QCD up to the third order in $\alpha_s(Q^2)$ \cite{Larin}:
\[
\displaystyle \hat{C}^{S}_1(Q^2) = 1 - 0.33333\Bigl({\frac{\alpha_s}{\pi}}\Bigr) - 0.54959\Bigl({\frac{\alpha_s}{\pi}}\Bigr)^2 - 4.44725\Bigl({\frac{\alpha_s}{\pi}}\Bigr)^3
\]
\[
\displaystyle C^{NS}_1(Q^2) = 1 - \Bigl({\frac{\alpha_s}{\pi}}\Bigr) - 3.5833\Bigl({\frac{\alpha_s}{\pi}}\Bigr)^2 - 20.2153\Bigl({\frac{\alpha_s}{\pi}}\Bigr)^3 .
\]
%~~~~~~~~~~~~~
With $\alpha_s$ evolved at the same order, one obtains
\begin{equation}
\hat{a}_0 = 0.33 \pm 0.03\ \mbox{(stat.)} \pm 0.05\ \mbox{(syst.)}.
\end{equation}
It should be noted here that the data have been evolved to
a common $Q^2$ on the basis of a fit at NLO only.
However, the choice of a value close to the average $Q^2$ of the data is expected to
minimise the effect of the evolution on the result quoted above.
Combining this value with $a_8$,
the first moment of the strange quark spin distribution in the limit $Q^2 \rightarrow \infty$
is found to be
\begin{equation}
  (\Delta s + \Delta{\overline s})_{Q^2\rightarrow\infty}
= \frac{1}{3} (\hat{a}_0 - a_8) = -0.08 \pm 0.01\ \mbox{(stat.)} \pm 0.02\ \mbox{(syst.)}.
\label{strange_Q2_inf}
\end{equation}
As stated before, this result relies on $SU(3)_f$ flavour symmetry. A 20\% symmetry
breaking, which is considered as a maximum \cite{a8}, would shift the value of
$\Delta s + \Delta{\overline s}$ by $\pm$ 0.04.

\begin{figure}
\epsfig{file=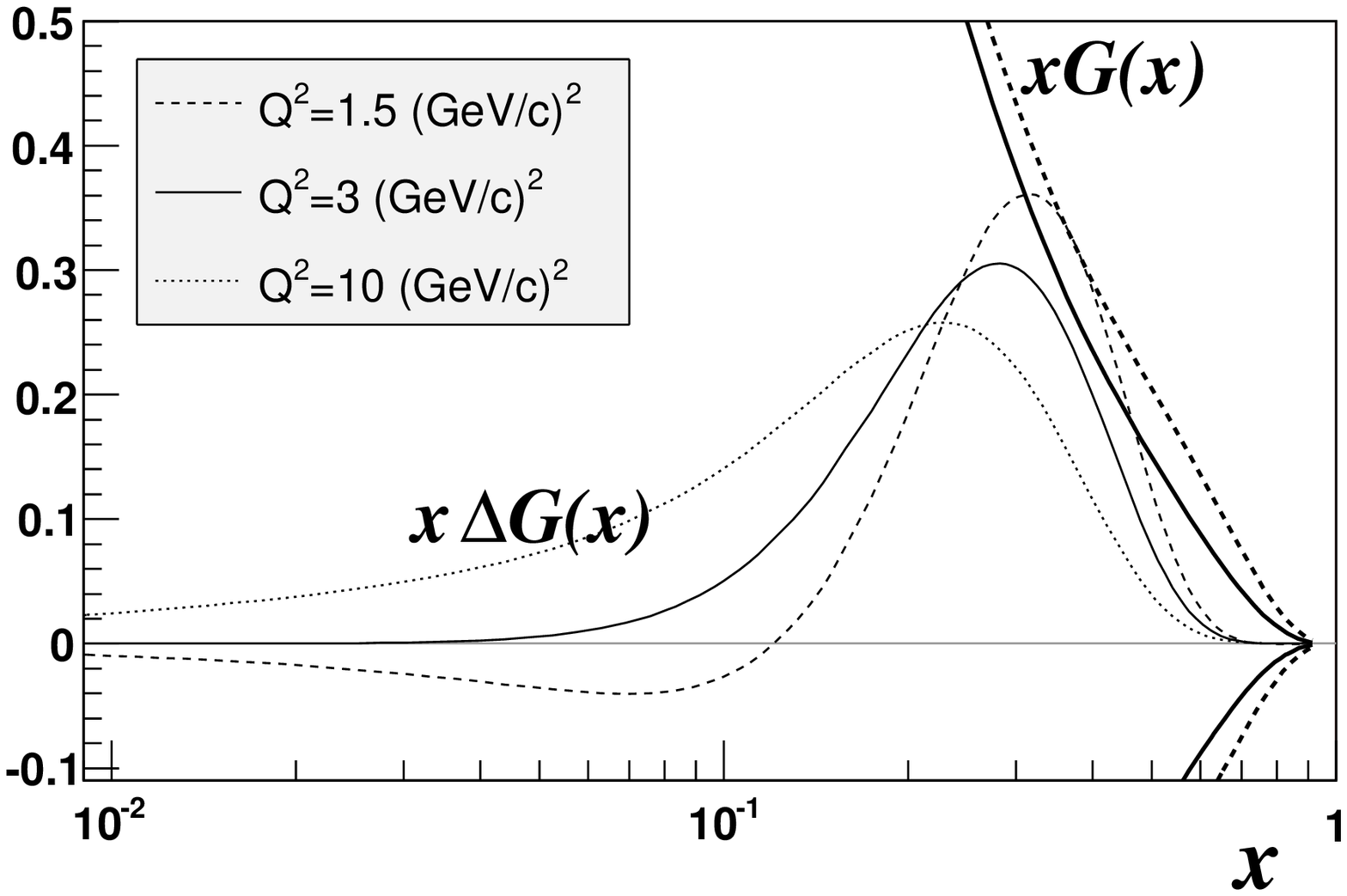,width=0.48\textwidth}
\hfill
\epsfig{file=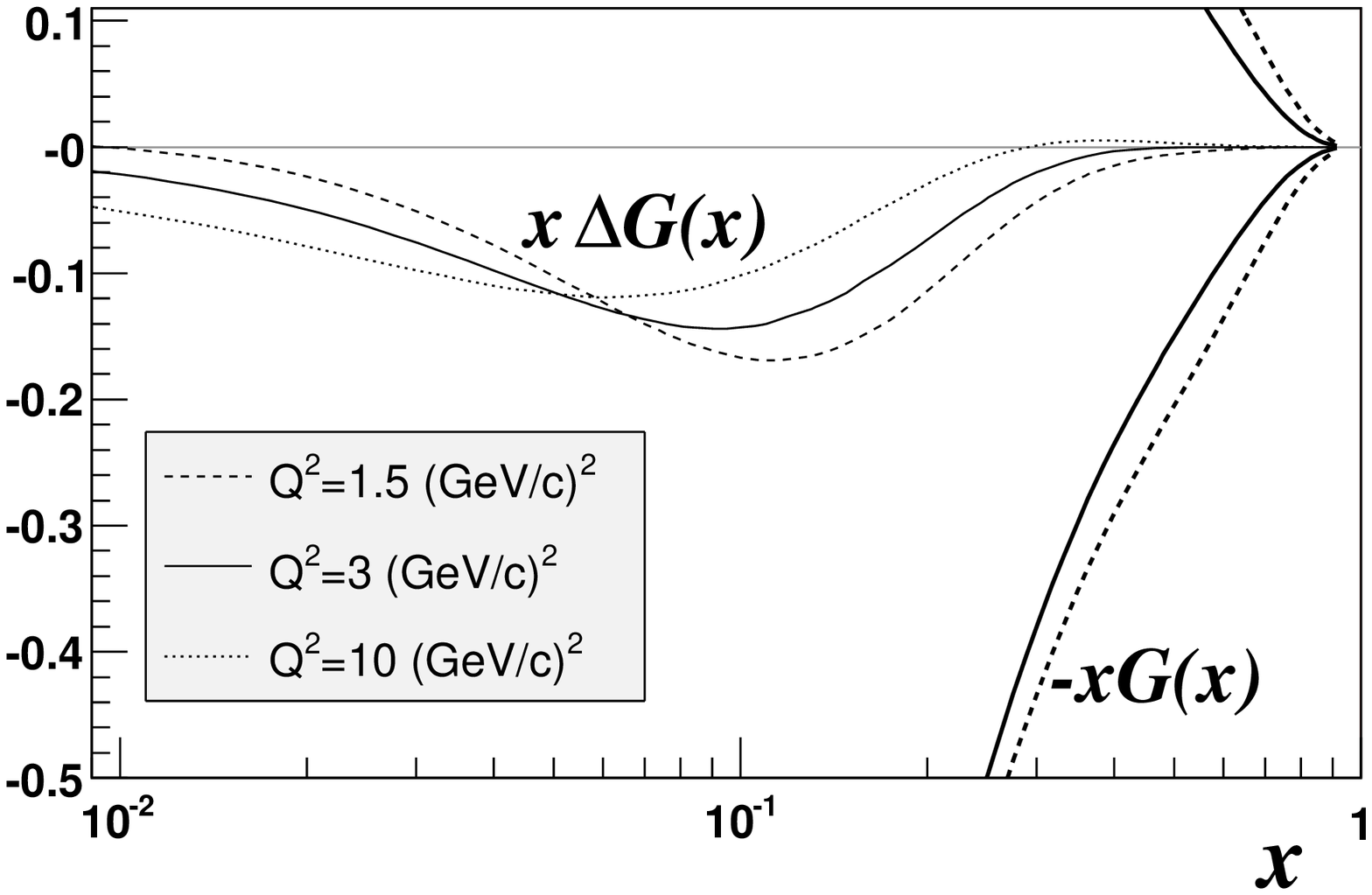,width=0.48\textwidth}
\caption{\small Gluon distribution $x \Delta G(x)$ corresponding to
  the fits with $\Delta G > 0$ ({left}) and $\Delta G < 0$ ({right}) obtained
  with the program of Ref.~\cite{smc_qcd}.
  The dashed, solid and dotted lines correspond to $Q^2 = 1.5$, 3 and 10~(GeV$/c)^2$,
  respectively.
  The unpolarised distributions $\pm x\, G(x)$ which were used in the fit
  as constrains for the polarised ones are shown for $Q^2 = 1.5$ and 3~(GeV$/c)^2$.}
\label{fig:Delta_G}
%\vspace{0.6cm}
\end{figure}

Previous fits of $g_1$, not including the COMPASS data, found a positive $\Delta G(x)$ and
a fitted function $g_1^d(x)$
%becoming negative below the first data point used at low $x$,
becoming negative for $x\lesssim0.025$ at $Q^2=3(\GeV/c)^2$,
as shown by the dotted line in Fig.~\ref{fig:g1_QCD_CMP_Q3}.
The new COMPASS data do not reveal any evidence for a decrease of the structure function at limit
%low $x$.
$x\rightarrow0$.
For our fit the data are still compatible with a positive $\Delta G$,
as shown by the full line in Fig.~\ref{fig:g1_QCD_CMP_Q3}.
%However in this case an unexpected dip appears at $x \simeq 0.25$. % in the fit with $\Delta G > 0$.
%at least with the usual parametrisation.
%A significant dip at $x \simeq 0.25$ is observed in
%$g_1^d(x,Q^2=3(\GeV/c)^2)$
%and becomes even stronger at lower $Q^2$.
However in this case a dip  at $x \simeq 0.25$ appears in the shape of
$g_1^d(x)$ for $Q^2 \rightarrow 1(\GeV/c)^2$.
Its origin  is related to the shape of the fitted $\Delta G(x)$,
%The origin of this dip is related to the shape of the fitted $\Delta G(x)$,
shown in Fig.~\ref{fig:Delta_G}\,(left).
Indeed, the gluon spin distribution must be close to zero at low $x$, to avoid pushing
$g_1^d$ down to negative values, and is also strongly limited at higher $x$ by the
positivity constraint $|\Delta G(x)| < G(x)$.
The whole distribution is thus squeezed in a narrow interval around the maximum
at $x \simeq \alpha_G/(\alpha_G + \beta_G)\simeq0.25$.

In contrast, the fit with negative $\Delta G$ reproduces very well the COMPASS low $x$
data with a much smoother distribution
of $\Delta G(x)$ (dashed line on Fig.~\ref{fig:g1_QCD_CMP_Q3})
and without approaching the positivity limit
(Fig.~\ref{fig:Delta_G}, right). The $(1 + \gamma x)$ factor in the singlet
quark distribution is not used in this case because it does not improve the confidence
level of the fit.

Comparing the fitted parameters for $\Delta G$ positive and negative (Table~\ref{tab:qcd_fit}),
we %first
observe that the parameters of the
non-singlet distributions $\Delta q_3(x)$ and $\Delta q_8(x)$ are
practically identical.
%very similar
The value of $\eta_{\Sigma}$ is
slightly larger
%by two $\sigma$'s larger
in the fit with $\Delta G < 0$, as could be expected since in this case $\Delta \Sigma(x)$
remains positive over the full range of $x$:
\begin{equation}
\eta_{\Sigma}\left(Q^2=3(\GeV/c)^2\right) = 0.27 \pm 0.01\ \mbox{(stat.)} ~ (\Delta G > 0),
\label{sigma_pl}
\end{equation}
\begin{equation}
\eta_{\Sigma}\left(Q^2=3(\GeV/c)^2\right) = 0.32 \pm 0.01\ \mbox{(stat.)} ~ (\Delta G < 0).
\label{sigma_min}
\end{equation}
We remind that in $\overline{\mbox{MS}}$ scheme $\eta_\Sigma$ is identical to the matrix element $a_0$.

%Taking the difference between the fits as an estimate of the systematic error,
%we obtain for the singlet moment derived from the fits to all $g_1$ data:
The singlet moment derived from the fits to all $g_1$ data is thus:
\begin{equation}
\eta_{\Sigma}\left(Q^2=3(\GeV/c)^2\right) = 0.30 \pm 0.01\ \mbox{(stat.)} \pm 0.02\ \mbox{(evol.)}.
\label{sigma_av}
\end{equation}
Here we have taken the difference between the fits as an estimate of the systematic error
and do not further investigate other contributions related to the choice of the QCD scale or the PDF
parametrisations.
The singlet moment obtained with COMPASS data alone (Eq.~(12)) is slightly above
this value and  its statistical error is larger by a factor of 3.
As stated before, the main uncertainty on the COMPASS result is due to the 10\%
normalisation uncertainty from the beam and target polarisation{s} and from the dilution factor.
The fact  that the COMPASS
data are on average slightly above  the world average can already be detected by a comparison
of the measured $g_1^d$ {values} to the
curves fitted to the world data (Fig.~\ref{fig:g1_QCD_CMP_Q3}).
{Hence} $a_0$ derived from the COMPASS value of $\Gamma_1^N$
is found to be slightly larger than $\eta_\Sigma$.

\begin{figure}
\epsfig{file=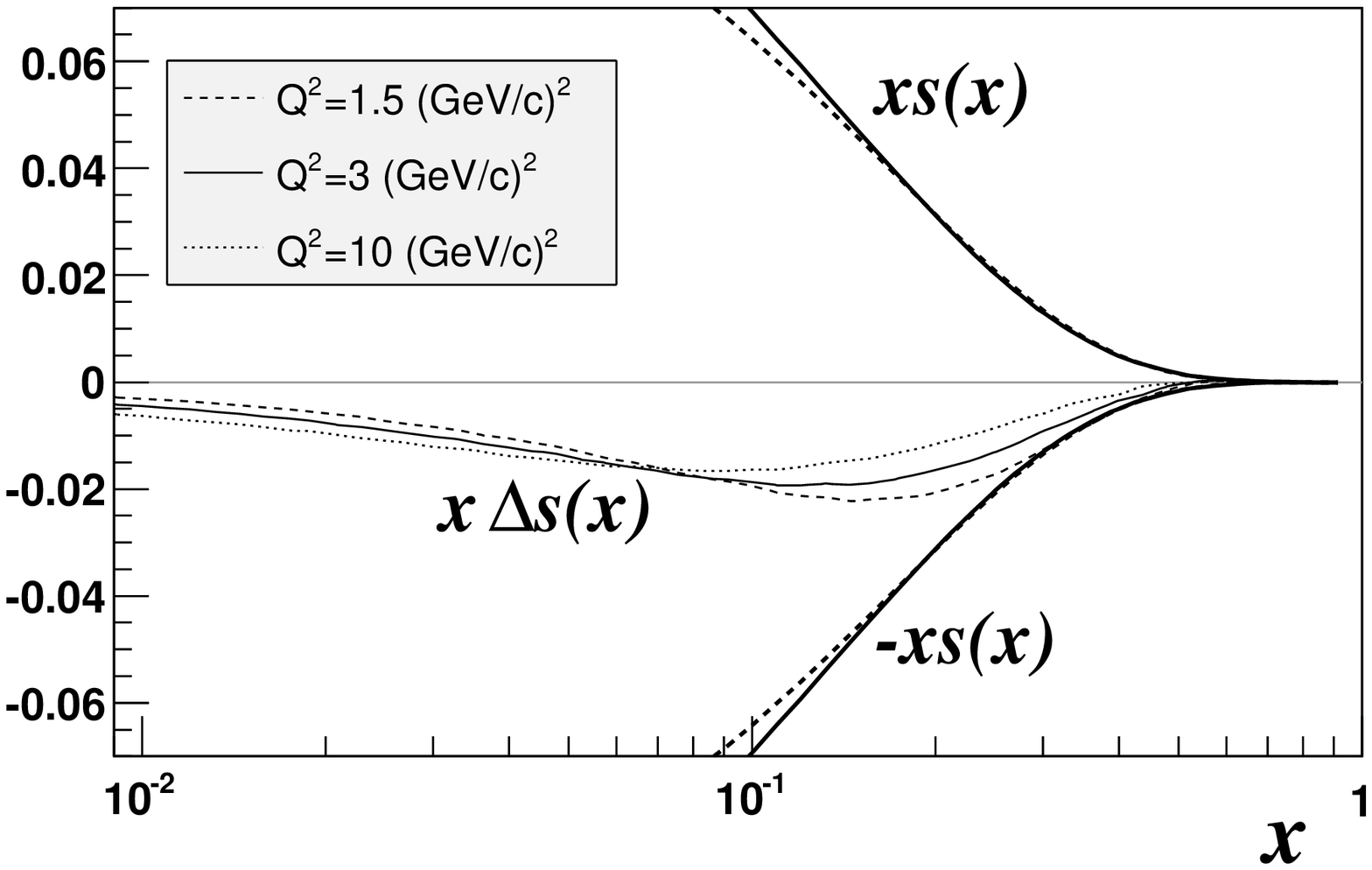,width=0.48\textwidth}
\hfill
\epsfig{file=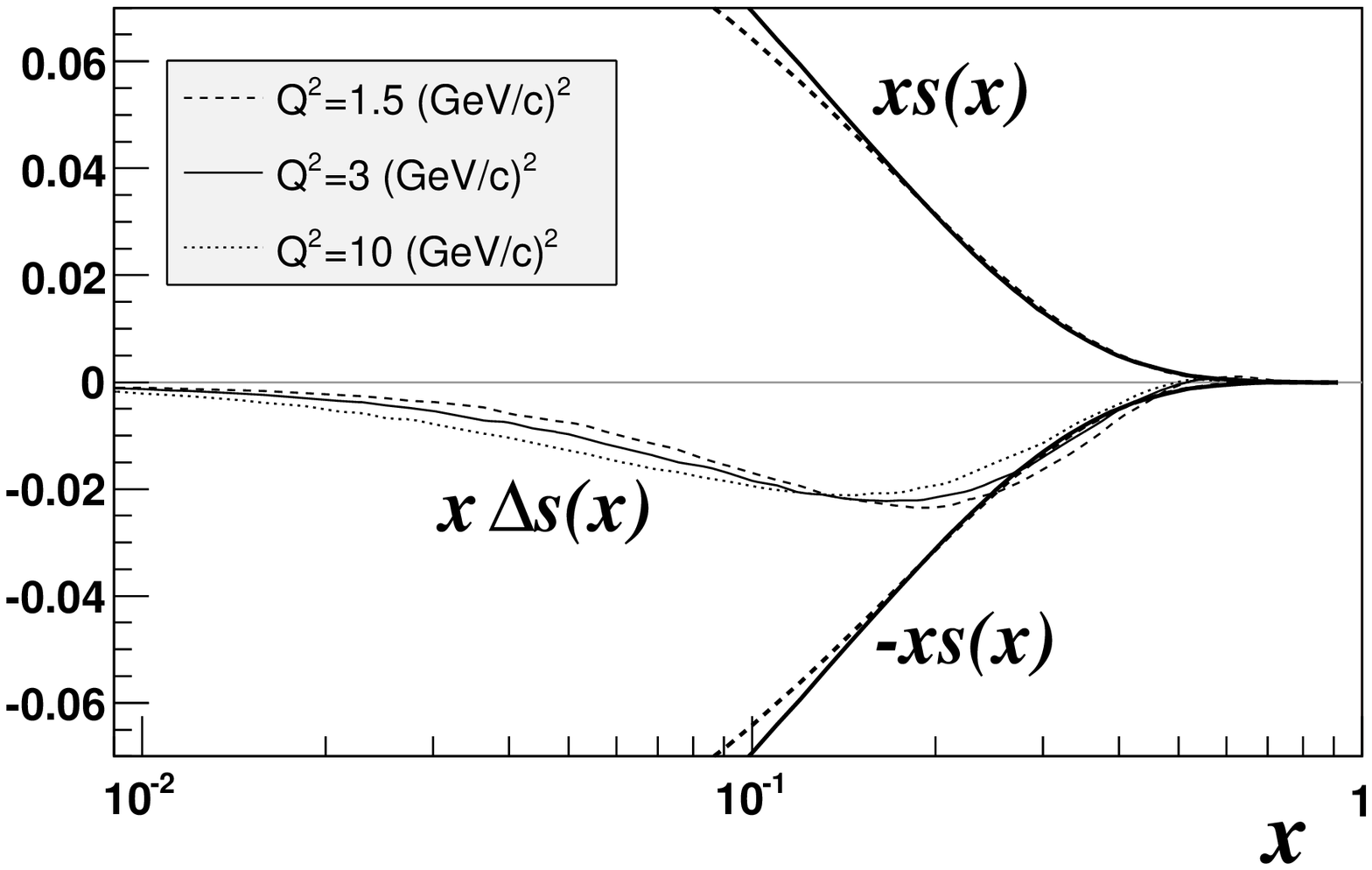,width=0.48\textwidth}
\caption{\small Strange quark  distribution $x \Delta s(x)$ corresponding
  to the fits with $\Delta G > 0$ ({left}) and $\Delta G < 0$ ({right}) obtained
  with the program of Ref.~\cite{smc_qcd}.
  The dashed, solid and dotted lines correspond to $Q^2 = 1.5$, 3 and 10~(GeV$/c)^2$,
  respectively. The unpolarised distributions $\pm x \, s(x)$ are shown for
  $Q^2 = 1.5$ and 3~(GeV$/c)^2$.}
\label{fig:Delta_s}
%\vspace{0.6cm}
\end{figure}

The polarised strange quark distributions, obtained from the difference between
$\Delta \Sigma(x)$ and $\Delta q_8(x)$ are shown in Fig.~\ref{fig:Delta_s}.
They are negative and concentrated in the highest $x$ region,
compatible with the constraint $|\Delta s(x)| < s(x)$.
This condition is indeed essential in the determination of the $\Delta q_8$
parameters which otherwise would be poorly constrained.
%It should be compared with the value of Eq.~(\ref{strange_Q2_inf}).}

\begin{figure}
\begin{center}
\epsfig{file=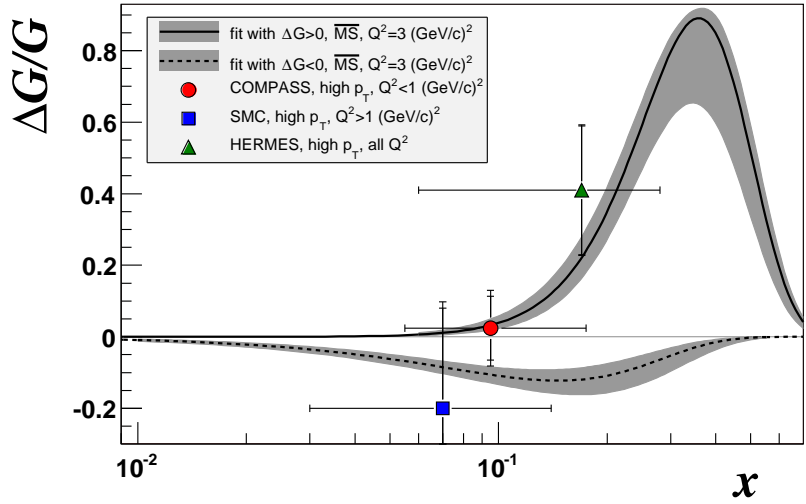,width=0.7\textwidth}
\end{center}
\caption{\small Distribution of the gluon polarisation $ \Delta G(x)/G(x)$
  at $Q^2=3(\GeV/c)^2$ for the fits with $\Delta G > 0$ and $\Delta G < 0$
  obtained with the program of Ref. \cite{smc_qcd}.
  The error bands correspond to the statistical error on $\Delta G(x)$ at a given $x$.
  The unpolarised gluon distribution is taken
  from the MRST parametrisation \cite{MRST}. The three data points show
  the measured values from SMC \cite{smc_hipt}, HERMES \cite{HERMES_hipt}
  and COMPASS \cite{cmp_hipt}.
  Two error bars are associated to each data point, one corresponding
  to the statistical precision and the other one to the statistical and systematic errors
  added in quadrature.
  The horizontal bar on each point shows the $x$-range of measurement.}
\label{fig:Delta_GbyG}
%\vspace{0.6cm}
\end{figure}

Although the gluon distributions
%of $\Delta G(x)$
strongly differ in the two fits,
the fitted values of
%$\eta_G$
their first moments are both small and about equal in absolute value
$|\eta_G|\approx0.2 - 0.3$.
We have also checked the stability of these results with respect to a
change in $\alpha_s(m_Z^2)$: when $\alpha_s(m_Z^2)$ is
varied by $\pm0.005$ the values of $\eta_G$ are not changed by more than
half a standard deviation.
%\begin{equation}
%\eta_G (Q^2=3\GeV^2) =  0.23~^{+~0.04}_{-~0.10}\ \mbox{(stat.)}~~~ (\Delta G > 0),
%\end{equation}
%and
%\begin{equation}
%\eta_G (Q^2=3\GeV^2) = -0.25~^{+~0.10}_{-~0.14}\ \mbox{(stat.)}~~~ (\Delta G < 0).
%\end{equation}
In Fig.~\ref{fig:Delta_GbyG} the existing direct measurements of $\Delta G/G$
\cite{smc_hipt,HERMES_hipt,cmp_hipt} are shown
with the distributions of $\Delta G(x)/G(x)$ derived from our fits
with $G(x)$ taken from Ref.~\cite{MRST}.
The HERMES value is positive and $2\sigma$ away from zero.
The measured SMC point is too unprecise to discriminate between positive or negative $\Delta G$.
The published COMPASS point, which has been obtained from a partial
data sample corresponding to about 40\% of the present statistics,
is almost on the $\Delta G > 0$ curve but is only $1.3 \sigma$ away from
the
%other
{$\Delta G < 0$} one,
so that no
%conclusion can be drawn so far.
{preference for any of the curves can be given so far}.
It should also be noted that the measured values of $\Delta G/G$ have
all been obtained in %an analysis at leading order in QCD.
{ leading order QCD analyses}.

In summary, we have measured the deuteron spin asymmetry $A_1^d$ and its longitudinal
spin-dependent structure function $g_1^d$ with improved precision at $Q^2>1(\GeV/c)^2$
over the range $0.004 < x < 0.70$.
The $g_1^d$ values are consistent with zero for $ x< 0.03$.
The measured values have been evolved to a common $Q^2$ by a new fit of the world $g_1$ data,
and the first moment $\Gamma_1^N$ has been evaluated at $Q^2 = 3(\GeV/c)^2$
with a statistical error smaller than 0.003.
From $\Gamma_1^N$ we have derived the  matrix element of the singlet
axial current $\hat{a}_0$ in the limit
$Q^2\rightarrow\infty$.
With COMPASS data alone, at the order $\alpha_s^3$, it has been found that
$\hat{a}_0 = 0.33 \pm 0.03\ \mbox{(stat.)} \pm 0.05\ \mbox{(syst.)}$
and the first moment of the strange quark distribution
$ (\Delta s + \Delta{\overline s})_{Q^2\rightarrow\infty} = -0.08 \pm 0.01\ \mbox{(stat.)} \pm 0.02\ \mbox{(syst.)}$.
We also observe that the fit
%\Blue{of the $Q^2$ evolution}
of world $g_1$ data at NLO yields two solutions with either $\Delta G(x) > 0$
or $\Delta G(x)<0$,
which equally well describe the present data.
In both cases, the first moment of $\Delta G(x)$ is of the order of 0.2--0.3 in absolute value
at $Q^2 = 3(\GeV/c)^2$ but the shapes of the distributions are very different.

\section*{Acknowledgements}
We gratefully acknowledge the support of the CERN management and staff and
the skill and effort of the technicians of our collaborating institutes.
Special thanks are due to V.~Anosov and V.~Pesaro for their technical support
during the installation and the running of this experiment.
This work was made possible by the financial support of our funding agencies.

%\newpage
\pagebreak

%\vspace{5mm}
\begin{table}
\begin{center}
\begin{tabular}{|c|c|c|r|r|}
\hline \hline
$x$ range & $\langle x \rangle$ & $\langle Q^2 \rangle $  & \multicolumn{1}{c|}{$A_1^d$} & \multicolumn{1}{c|}{$g_1^d$} \\
 &  & [(GeV$/c)^2]$  &  &  \\
\hline \hline
 0.0030--0.0035 & 0.0033 & ~0.78 & $ 0.003 \pm 0.009 \pm 0.004$ & $ 0.090 \pm 0.240 \pm 0.107$ \\
 0.0035--0.0040 & 0.0038 & ~0.83 & $-0.004 \pm 0.007 \pm 0.003$ & $-0.097 \pm 0.183 \pm 0.082$ \\
\hline \hline
% 0.004--0.005 & 0.0046 & ~1.10 & $ 0.004 \pm 0.009 \pm 0.004$ & $ 0.082 \pm 0.210 \pm 0.089$ \\
% 0.005--0.006 & 0.0055 & ~1.22 & $ 0.003 \pm 0.007 \pm 0.003$ & $ 0.062 \pm 0.146 \pm 0.062$ \\
% 0.006--0.008 & 0.0070 & ~1.39 & $-0.002 \pm 0.005 \pm 0.002$ & $-0.034 \pm 0.086 \pm 0.036$ \\
% 0.008--0.010 & 0.0090 & ~1.61 & $-0.010 \pm 0.006 \pm 0.003$ & $-0.139 \pm 0.078 \pm 0.035$ \\
% 0.010--0.020 & 0.0141 & ~2.15 & $ 0.002 \pm 0.004 \pm 0.002$ & $ 0.017 \pm 0.033 \pm 0.014$ \\
% 0.020--0.030 & 0.0244 & ~3.18 & $ 0.003 \pm 0.006 \pm 0.003$ & $ 0.017 \pm 0.035 \pm 0.015$ \\
% 0.030--0.040 & 0.0346 & ~4.26 & $ 0.009 \pm 0.008 \pm 0.004$ & $ 0.041 \pm 0.035 \pm 0.016$ \\
% 0.040--0.060 & 0.0487 & ~5.80 & $ 0.017 \pm 0.008 \pm 0.004$ & $ 0.054 \pm 0.026 \pm 0.012$ \\
% 0.060--0.100 & 0.0765 & ~8.53 & $ 0.058 \pm 0.009 \pm 0.007$ & $ 0.121 \pm 0.019 \pm 0.014$ \\
% 0.100--0.150 & 0.1212 & 12.67 & $ 0.095 \pm 0.013 \pm 0.011$ & $ 0.123 \pm 0.017 \pm 0.014$ \\
% 0.150--0.200 & 0.1719 & 17.23 & $ 0.123 \pm 0.020 \pm 0.014$ & $ 0.103 \pm 0.016 \pm 0.012$ \\
% 0.200--0.250 & 0.2223 & 21.86 & $ 0.183 \pm 0.028 \pm 0.021$ & $ 0.106 \pm 0.016 \pm 0.012$ \\
% 0.250--0.350 & 0.2905 & 28.38 & $ 0.216 \pm 0.030 \pm 0.024$ & $ 0.077 \pm 0.011 \pm 0.009$ \\
% 0.350--0.500 & 0.4050 & 39.77 & $ 0.343 \pm 0.049 \pm 0.038$ & $ 0.055 \pm 0.008 \pm 0.006$ \\
% 0.500--0.700 & 0.5660 & 55.30 & $ 0.626 \pm 0.112 \pm 0.075$ & $ 0.027 \pm 0.005 \pm 0.003$ \\
 0.004--0.005 & 0.0046 & ~1.10 & $ 0.004 \pm 0.009 \pm 0.004$ & $ 0.082 \pm 0.210 \pm 0.089$ \\
 0.005--0.006 & 0.0055 & ~1.22 & $ 0.003 \pm 0.007 \pm 0.003$ & $ 0.062 \pm 0.146 \pm 0.062$ \\
 0.006--0.008 & 0.0070 & ~1.39 & $-0.002 \pm 0.005 \pm 0.002$ & $-0.034 \pm 0.086 \pm 0.036$ \\
 0.008--0.010 & 0.0090 & ~1.61 & $-0.010 \pm 0.006 \pm 0.003$ & $-0.139 \pm 0.078 \pm 0.035$ \\
 0.010--0.020 & 0.0141 & ~2.15 & $ 0.002 \pm 0.004 \pm 0.002$ & $ 0.017 \pm 0.033 \pm 0.014$ \\
 0.020--0.030 & 0.0244 & ~3.18 & $ 0.003 \pm 0.006 \pm 0.003$ & $ 0.017 \pm 0.035 \pm 0.015$ \\
 0.030--0.040 & 0.0346 & ~4.26 & $ 0.009 \pm 0.008 \pm 0.004$ & $ 0.041 \pm 0.035 \pm 0.016$ \\
 0.040--0.060 & 0.0487 & ~5.80 & $ 0.017 \pm 0.008 \pm 0.004$ & $ 0.054 \pm 0.026 \pm 0.012$ \\
 0.060--0.100 & 0.0765 & ~8.53 & $ 0.058 \pm 0.009 \pm 0.007$ & $ 0.121 \pm 0.019 \pm 0.014$ \\
 0.100--0.150 & 0.121~\, & 12.6~~ & $ 0.095 \pm 0.013 \pm 0.011$ & $ 0.123 \pm 0.017 \pm 0.014$ \\
 0.150--0.200 & 0.171~\, & 17.2~~ & $ 0.123 \pm 0.020 \pm 0.014$ & $ 0.103 \pm 0.016 \pm 0.012$ \\
 0.200--0.250 & 0.222~\, & 21.8~~ & $ 0.183 \pm 0.028 \pm 0.021$ & $ 0.106 \pm 0.016 \pm 0.012$ \\
 0.250--0.350 & 0.290~\, & 28.3~~ & $ 0.216 \pm 0.030 \pm 0.024$ & $ 0.077 \pm 0.011 \pm 0.009$ \\
 0.350--0.500 & 0.405~\, & 39.7~~ & $ 0.343 \pm 0.049 \pm 0.038$ & $ 0.055 \pm 0.008 \pm 0.006$ \\
 0.500--0.700 & 0.566~\, & 55.3~~ & $ 0.626 \pm 0.112 \pm 0.075$ & $ 0.027 \pm 0.005 \pm 0.003$ \\
\hline \hline
\end{tabular}
\caption{\small
  Values of $A_1^d$ and $g_1^d$ with their statistical and systematical errors
  as a function of $x$ with the corresponding average values of $x$ and $Q^2$.
  The minimum $Q^2$ cut is 1~(GeV$/c)^2$ except for the first two points where
  it is lowered to 0.7~(GeV$/c)^2$. These two data points are shown on the figures
  as complementary information but were not used in the fits.}
\label{tab:a1_g1}
\end{center}
\end{table}

\begin{table}
\begin{center}

\begin{tabular}{|l|l|c|c|}
\hline
\hline
              & Beam polarization     & $dP_B/P_B$ & 5\% \\
\cline{2-4}
Multiplicative & Target polarization   & $dP_T/P_T$ & 5\% \\
\cline{2-4}
variables      & Depolarization factor & $d D(R)/D(R)$ & 2 -- 3 \% \\
\cline{2-4}
error, $\Delta A_1^{mult}$& Dilution factor & $df/f$    & 6 \% \\
\cline{2-4}
\cline{2-4}
              & Total             &   & $\Delta A_1^{mult} \simeq 0.1 A_1$ \\
\hline
\hline
Additive       & Transverse asymmetry  & ${\eta}/\rho\cdot\Delta A_2$ & $10^{-4}-5\cdot10^{-3}$ \\
\cline{2-4}
variables      & Radiative corrections& $\Delta A_1^{RC}$ &
$10^{-4}-10^{-3}$\\
\cline{2-4}
error, $\Delta A_1^{add}$ & False asymmetry   & $A_{false}$ & $<0.4\cdot \Delta A_1^{stat}$ \\
\hline
\hline
\end{tabular}
\end{center}
\caption{\small Decomposition of the systematic error of $A_1$
into multiplicative and additive variables contributions.}
\label{tab:sys_error}
\end{table}

\begin{table}
%\begin{center}
\begin{minipage}{0.5\linewidth}
%\begin{center}

\noindent
\begin{tabular}{|c|r|r|}
\hline  \hline
\multicolumn{3}{|c|}
%  { Fit~~1 ~~~~~~~~~ $\Delta G > 0$} \\
{$\Delta G > 0$} \\
\hline
                   &  Prog. Ref. \cite{smc_qcd}      & Prog. Ref. \cite{Sissakian}     \\
\hline
 $\eta_{\Sigma}$   &  $0.270 \pm  0.014$             & $0.284~^{+~0.016}_{-~0.014}$~~  \\
 $\alpha_{\Sigma}$ &$-0.303~^{+~0.074}_{-~0.079}$~~~ & $-0.226~^{+~0.103}_{-~0.101}$~~\\
 $\beta_{\Sigma}$  &  $3.60 ^{+~0.24}_{-~0.22}$~~~   & $3.69~^{+~0.30}_{-~0.25}$~~~   \\
 $\gamma_{\Sigma}$ & $-16.0~^{+~1.4}_{-~1.6}$~~~~    & $-15.8~^{+~1.7}_{-~2.8}$~~~~   \\
\hline
 $\eta_G$          & $0.336 ^{+~0.049}_{-~0.070}$~~  & $0.233~^{+~0.040}_{-~0.053}$~~ \\
 $\alpha_G$        & $2.91 ^{+~0.40}_{-~0.44}$~~~    & $3.11^{+~0.42}_{-~0.53}$~~~     \\
 $\beta_G$         & $10$ (fixed)~~~~                & $10$  (fixed)~~~~                \\
\hline
 $\alpha_3$        &  $-0.226 \pm 0.027$             & $-0.226~^{+~0.029}_{-~0.027}$~~\\
 $\beta_3 $        & $2.43 ^{+~0.11}_{-~0.10}$~~~    & $2.38~^{+~0.11}_{-~0.10}$~~~   \\
\hline
 $\alpha_8$        & $0.35 ^{+~0.18}_{-~0.44}$~~~    & $0.45~^{+~0.13}_{-~0.43}$~~~   \\
 $\beta_8$         & $3.36 ^{+~0.60}_{-~1.04}$~~~    & $3.50~^{+~0.46}_{-~0.98}$~~~   \\
\hline
$\chi^2/{\rm ndf}$ &  233/219~~~~                    &  232/219~~~~                   \\
\hline  \hline
\end{tabular}
%\end{center}
\end{minipage}
\hfill
\begin{minipage}{0.5\linewidth}
%\begin{center}

\noindent
\begin{tabular}{|c|r|r|}
\hline  \hline
\multicolumn{3}{|c|}
%   {Fit~~2 ~~~~~~~~~$\Delta G < 0 $} \\
{$\Delta G < 0 $} \\
\hline
                   &  Prog. Ref. \cite{smc_qcd}    & Prog. Ref. \cite{Sissakian}  \\
\hline
 $\eta_{\Sigma}$   &  $0.320 \pm 0.009$            &  $0.328 \pm 0.009$           \\
 $\alpha_{\Sigma}$ &  $1.38  ^{+~0.15}_{-~0.14}$~~~& $1.38~^{+~0.13}_{-~0.12}$~ \\
 $\beta_{\Sigma}$  &  $4.08  ^{+~0.29}_{-~0.27}$~~~& $4.05~^{+~0.25}_{-~0.23}$~ \\
 $\gamma_{\Sigma}$ &   \multicolumn{1}{c|}{-}      & \multicolumn{1}{c|}{-}       \\
\hline
 $\eta_G$          &$-0.309^{+~0.095}_{-~0.144}$~~~& $-0.192~^{+~0.064}_{-~0.109}$ \\
 $\alpha_G$        & $0.39 ^{+~0.65}_{-~0.48}$~~~  & $0.23~^{+0.063}_{-~0.47}$  \\
 $\beta_G$         &   $13.9 ^{+~7.8}_{-~5.4}$~~~~ & $13.8~^{+~8.2}_{-~5.6}$~~~ \\
\hline
$\alpha_3$         &$-0.212 \pm 0.027$             & $-0.209   \pm 0.027$        \\
$\beta_3 $         & $2.44 ^{+~0.11}_{-~0.10}$~~~  & $2.40~^{+~0.11}_{~-0.10}$~~ \\
\hline
$\alpha_8$         & $0.43~^{+~0.15}_{-~0.16}$     & $0.383~^{+~0.080}_{-~0.121}$ \\
$\beta_8$          & $3.54 ^{+~0.55}_{-~0.54}$~~~  & $3.39~^{+~0.33}_{-~0.39}$~ \\
\hline
$\chi^2/{\rm ndf}$ &  247/219~~~~                  &  247/219~~~~        \\
\hline  \hline
\end{tabular}
\end{minipage}

\vspace*{1cm}

\begin{center}
\def\diag{}
\def\fix{\multicolumn{1}{c}{--}}
\def\fie{\multicolumn{1}{c|}{--}}
\scriptsize{
\begin{tabular}{|c|rrrrrrrrrrr|}
\hline  \hline
&\multicolumn{1}{c}{ $\eta_{\Sigma}$ }
&\multicolumn{1}{c}{ $ \alpha_{\Sigma}$ }
&\multicolumn{1}{c}{ $\beta_{\Sigma}$ }
&\multicolumn{1}{c}{ $\gamma_{\Sigma}$}
&\multicolumn{1}{c}{ $\eta_G$}
&\multicolumn{1}{c}{ $\alpha_G$ }
&\multicolumn{1}{c}{ $\beta_G$ }
&\multicolumn{1}{c}{ $\alpha_3$}
&\multicolumn{1}{c}{ $\beta_3 $}
&\multicolumn{1}{c}{ $\alpha_8 $}
&\multicolumn{1}{c|}{ $\beta_8$ }\\
\hline
$\eta_{\Sigma}  $&  \diag &$ 0.581$&$ 0.143$&$-0.432$&$-0.548$&$ 0.549$&  \fix  &$-0.075$&$-0.118$&$ 0.030$&$-0.008$\\
$\alpha_{\Sigma}$&$-0.492$&  \diag &$ 0.648$&$ 0.272$&$-0.434$&$ 0.452$&  \fix  &$ 0.053$&$ 0.066$&$-0.121$&$-0.047$\\
$\beta_{\Sigma} $&$-0.388$&$ 0.877$&  \diag &$ 0.304$&$-0.011$&$ 0.022$&  \fix  &$-0.010$&$-0.037$&$-0.420$&$-0.499$\\
$\gamma_{\Sigma}$&  \fix  &  \fix  &  \fix  &  \diag &$ 0.272$&$-0.248$&  \fix  &$ 0.088$&$ 0.142$&$-0.361$&$-0.025$\\
$\eta_G         $&$ 0.277$&$-0.221$&$-0.130$&  \fix  &  \diag &$-0.978$&  \fix  &$ 0.082$&$ 0.066$&$ 0.071$&$ 0.067$\\
$\alpha_G       $&$ 0.162$&$-0.052$&$ 0.012$&  \fix  &$ 0.835$&  \diag &  \fix  &$-0.087$&$-0.070$&$-0.069$&$-0.063$\\
$\beta_G        $&$ 0.148$&$-0.039$&$ 0.025$&  \fix  &$ 0.814$&$ 0.935$&  \diag &  \fix  &  \fix  & \fix   & \fie   \\
$\alpha_3       $&$-0.012$&$ 0.008$&$-0.032$&  \fix  &$ 0.078$&$ 0.006$&$ 0.053$&  \diag &$ 0.788$&$-0.023$&$-0.020$\\
$\beta_3        $&$-0.104$&$ 0.067$&$ 0.037$&  \fix  &$ 0.060$&$ 0.003$&$ 0.023$&$ 0.793$&  \diag &$-0.017$&$-0.013$\\
$\alpha_8       $&$-0.105$&$-0.175$&$-0.276$&  \fix  &$ 0.171$&$ 0.099$&$ 0.219$&$-0.036$&$-0.016$&  \diag &$ 0.832$\\
$\beta_8        $&$-0.137$&$ 0.033$&$-0.211$&  \fix  &$ 0.118$&$ 0.063$&$ 0.138$&$-0.044$&$-0.026$&$ 0.821$&  \diag \\
\hline \hline
\end{tabular}
}
\end{center}
\psset{linewidth=0.5pt}
\qline(1.20,3.92)(15.65,0.25)

\caption{\small Top: Values of the parameters obtained from the QCD analysis
  at $Q^2 = 3$ (GeV$/c)^2$ in fits
  with $\Delta G > 0$ and $\Delta G < 0$ with the two programs.
  The quoted errors correspond to one $\sigma$ and have been obtained
  from the MINOS analysis \cite{minuit}.
  The strongly asymmetric errors obtained for some parameters are due to
  the positivity constraints applied in the fits.
  Bottom: Correlation matrices for the fits by the program of Ref.~\cite{smc_qcd}.
  The triangles above and below the diagonal correspond to the fits with $\Delta G > 0$
  and $\Delta G < 0$, respectively. The ``--" symbols correspond to parameters which
  are fixed in one of the fits.}

\label{tab:qcd_fit}
%\end{center}
\end{table}

\begin{table}

\begin{center}
\begin{tabular}{|c||c|c||c|c|}
\hline
& \multicolumn{4}{c|}{COMPASS data evolved to $Q^2=3(\GeV/c)^2$ using} \\
\cline{2-5}
Range in $x$ & \multicolumn{2}{c||}{fits of}  & \multicolumn{2}{c|}{COMPASS fits (prog. \cite{smc_qcd})} \\
\cline{2-5}
& ~~~~BB{\cite{BB}}~~ & ~~LSS{\cite{LSS05}}~~ & ~~$\Delta G>0$~~ & ~~$\Delta G<0$~~ \\
\hline
\hline
$[\,0.004,\,0.7\,]$& ~~$0.0455$  & ~~$0.0469$  & ~~$0.0469$  & $0.0511$ \\
\hline
$[\,0.7,\,1\,]$    & ~~$0.0014$  & ~~$0.0008$  & ~~$0.0011$  & $0.0010$  \\
\hline
$[\,0,\,0.004\,]$  & $-0.0040$ & $-0.0029$ & $-0.0014$ & $0.0004$  \\
\hline
\hline
$[\,0,\,1\,]$      & ~~$0.0430$  &  ~~$0.0448$ & ~~$0.0466$  & $0.0525$ \\
\hline
\end{tabular}
\caption{\small Contributions to $\Gamma_1^N\left(Q^2=3(\GeV/c)^2\right)$
  from different kinematic regions.
  The values in the first line are the COMPASS results evolved according
  to different fits and integrated over the measured $x$ range.
  The second and third lines show the corresponding high and low $x$ extrapolations.}
\label{tab:Gamma_1}
\end{center}
\end{table}

%%%%%%%%%%%%%%%%%%%%%%%%%%%%%%%%%%%%%%%%%%%%%%%%%%%%%%%%%%%%%%%%%%
%\newpage
\pagebreak

%\begin{figure}[]
%  \begin{center}
%    \includegraphics[width=0.29\textwidth,clip]{frac_x_020304.eps}
%    \includegraphics[width=0.29\textwidth,clip]{frac_Q2_020304.eps}
%    \includegraphics[width=0.29\textwidth,clip]{test_ct.eps}
%    \caption{\small Fraction of inclusive, semi-inclusive, and calorimetric triggers
%      as a function of $x$ (left) and $Q^2$ (center).
%      Events are counted with the weight they carry in the asymmetry calculation.
%      Difference between asymmetries for inclusive and hadronic events in the
%      kinematic range covered by the purely calorimetric trigger (right).}
%\label{fig:trigg_xq2}
%  \end{center}
%\end{figure}

\end{document}